# EXPLORING CHALLENGES AND OPPORTUNITIES IN CYBERSECURITY RISK AND THREAT COMMUNICATIONS RELATED TO THE MEDICAL INTERNET OF THINGS (MIOT)

George W. Jackson, Jr.[1] and Shawon S. M. Rahman[2]

[1]Capella University, Minneapolis, MN, USA
[2]Associate Professor, Department of Computer Science and Engineering
University of Hawaii-Hilo, Hilo, Hawaii, USA

## *ABSTRACT*

*As device interconnectivity and ubiquitous computing continues to proliferate healthcare, the Medical Internet of Things (MIoT), also well known as the, Internet of Medical Things (IoMT) or the Internet of Healthcare Things (IoHT), is certain to play a major role in the health, and well-being of billions of people across the globe. When it comes to issues of cybersecurity risks and threats connected to the IoT in all of its various flavors the emphasis has been on technical challenges and technical solution. However, especially in the area of healthcare, there is another substantial and potentially grave challenge. It is the challenge of thoroughly and accurately communicating the nature and extent of cybersecurity risks and threats to patients who are reliant upon these interconnected healthcare technologies to improve and even preserve their lives. This case study was conducted to assess the scope and depth of cybersecurity risk and threat communications delivered to an extremely vulnerable patient population, semi-structured interviews were held with cardiac medical device specialists across the United States. This research contributes to scientific data in the field of healthcare cybersecurity and assists scholars and practitioners in advancing education and research in the field of MIoT patient communications.*

## *KEYWORDS*

*Internet of Things, IoT Security, Medical Internet of Things, Healthcare Cybersecurity, Thematic Analysis*

## 1. INTRODUCTION

As device interconnectivity and ubiquitous computing continues to proliferate healthcare the Internet of Things (MIoT), also well known as the, Internet of Medical Things (IoMT) or the Internet of Healthcare Things (IoHT), is certain to play a major role in the health, and well-being of billions of people across the globe. Because of the ubiquity of internetworking technologies, such as the MIoT, wireless implanted medical devices are experiencing dramatic growth in their usage, application, and complexity [1-5]. This rapid growth in usage and complexity brings the increased threat of cybersecurity attacks the same as any internet-enabled device might experience [2, 3, 5, 6, 7, 8, 28].

This study examined the question: How are the cybersecurity risks and threats related to wireless implanted medical devices being communicated to patients who have or will have these devices implanted in their bodies? By using semi-structured interviews with expert cardiac device specialists, this multiple case study examined the ideas, experiences, and perceptions of the healthcare professionals directly responsible for the cybersecurity [30] awareness and education of an extremely high-profile patient population. Thematic Analysis was used to scrutinize and categorize the con-tent of these frank and in-depth conversations on the current state of communications with patients reliant upon interconnected medical device technology.

Through the application of Thematic Analysis (TA), this multiple case study yielded five themes, related to the issues and challenges related to cybersecurity risk [29] and threat communications with MIoT patients. Those themes were: the influence of the media on patient cybersecurity awareness; the need for risk/benefit analysis at all levels of patient interaction; the culture of non-communication in the healthcare field and medical device industry; the need for collaboration and education among manufacturers, healthcare providers and patients; and the obstacles and challenges present in all phases of MIoT patient care. To provide additional context to the case study described herein this paper includes a brief literature review of the MIoT as it exists today and its potential future directions. This combined with the information gleaned from healthcare professionals working closest with the most vulnerable MIoT patient population (namely, cardiac device patients) provides valuable insights to medical device manufacturers, healthcare providers, policymakers and patients across the interconnected medical device macrocosm.

When it comes to issues of cybersecurity risks and threats connected to the internet of things (IoT) [25] in all its various flavors (for example automotive, retail, industrial, smart video, energy, smart buildings, and healthcare) the emphasis has been on addressing technical challenges and the delivery of technical solutions. However, especially in healthcare, there is another substantial and potentially grave challenge to be addressed. That challenge is the challenge of thoroughly and accurately communicating the nature and extent of cybersecurity risks and threats to patients who are reliant upon these interconnected healthcare technologies to improve and even preserve their lives.

In recent multiple case studies conducted to assess the scope and depth of cybersecurity risk [26] and threat communications delivered to an extremely vulnerable patient population, semi-structured interviews were held with cardiac medical device specialists across the United States. These cardiac medical devices experts were intimately involved in the care and patient safety of cardiac implantable electronic devices (CIED) patients who used home monitoring systems to connect their implantable cardioverter-defibrillator (ICD) and pacemaker (PM) devices, typically through the cloud[23], to their physician's offices or device clinic. The case study examined the question: How are the cybersecurity risks[30] and threats related to wireless implanted medical devices being communicated to patients who have or will have these devices implanted in their bodies? The subjects of this multiple case study were sixteen cardiac device specialists in the U.S., each possessing at least one year of experience working directly with CIED patients, who actively used cardiac device home monitoring systems.

By using semi-structured interviews with expert implanted cardiac device specialists, this multiple case study examined the ideas, experiences, and perceptions of the healthcare professionals directly responsible for the cybersecurity awareness and education of CIED patients in an extremely high-profile patient population. Through the application of Thematic Analysis (TA), this multiple case study yielded five themes, related to the research question. Those themes were: the influence of the media on patient cybersecurity awareness; the need for risk/benefit analysis at all levels of patient interaction; the culture of non-communication in the healthcare field and medical device industry; the need for collaboration and education among manufacturers, healthcare providers and patients; and the obstacles and challenges present in all phases of patient care.

To provide additional context to the case study described later in this paper, the following section provides a brief literature review of the MIoT as it exists today and its potential future directions.

This combined with the information gleaned from healthcare professionals working closest with the most vulnerable MIoT patient population (namely, cardiac device patients) provides valuable insights to medical device manufacturers, healthcare providers, policymakers and patients across the interconnected medical device macrocosm.

## 2. LITERATURE REVIEW

The focus of [9] was the security and privacy of the data flow within the MIoT infrastructure. In the course of its analysis [9] gave a brief overview of the MIoT. According to [9]: "Generally, the MIoT structure is composed of three layers: the perception layer, the network layer, and the application layer, as shown in Figure 1. The major task of the perception layer is to collect healthcare data with a variety of devices. The network layer, which is composed of wired and wireless system and middleware, processes and transmits the input obtained by the perception layer supported by technological platforms. Well-designed transport protocols not only improve transmission efficiency and reduce energy consumption but also ensure security and privacy. The application layer integrates the medical information resources to provide personalized medical services and satisfy the final users' needs, according to the actual situation of the target population and the service demand."

A key driver of the MIoT is the ability to monitor patients remotely. Healthcare providers and patients enjoy the benefits of having their vital health-related data and statistics monitored remotely. Not only does this save patients and providers time and expense but the interconnectivity of the MIoT can provide continuous, real-time monitoring and data processing. Other aspects adding to the desirability of interconnected medical devices is that devices can share data through the cloud with other medical applications such as an electronic medical record (EMR). The proliferation of cloud services[31] and cloud storage adds to the convenience and efficiency of delivering healthcare while providing opportunities to drive down costs. These steadily increasing benefits, however, come with steadily increasing risks. As the MIoT grows the amount of data that flows through these interconnected devices increases dramatically. The more interconnected medical devices there are more, points of failure arise. What is more the increase and expansion of networking communications increase [27] the attack surface for cybercriminals and malicious actors resulting in the increased risks and threats to sensitive data. Various technical solutions were discussed for MIoT security and privacy such in [9] such as data encryption schemes, access control mechanisms, trusted third party auditing, and data anonymization methods. However, the study concluded that most MIoT implementation continues to face challenges with insecure networks, limitations of power, storage and memory capacity within an individual or collective components of the MIoT infrastructure which render significant portions of the MIoT system vulnerable to cyberattack. Nevertheless, [9] asserts that while much attention has been given to the MIoT and great improvements have been made in MIoT privacy and security there is still a great deal of work and research to be done.

 In [10] the Internet of Health Things (IoHT) was the name used for the internetworking of medical devices. Once again [10] focused on the technical challenges of the IoHT and future trends.

The study defines the Internet of Health Things (IoHT) as an IoT-based solution that includes the ability to connect patients to healthcare facilities [10]. Like [9], [10] highlights the positive impact remote access to patient's data has upon healthcare convenience, efficiency and cost reduction. Likewise, both [9] and [10] point out that smartphones will play an ever-increasing role in the healthcare IoT space as the mobile phone[24] is undoubtedly the most ubiquitous of all networking devices. Rodrigues et al., (2018) also gave a great deal of attention to wearable devices as are of the IoHT. The study concluded that healthcare has been one of the fastest industries to embrace IoT. This has led to a growing trend of established companies and start-up worldwide moving into the interconnected healthcare device arena.

## 3. Case Study: Complex Communication Challenges Within The MIoT

Once a medical device whether it be standalone, wearable or implanted is has become networked and interconnected with other medical devices it has become part of the MIoT. Combating MIoT cybersecurity threats is an incredibly complex undertaking that requires a concerted effort across multiple layers of stakeholders, e.g., device manufacturers, devices suppliers, government agencies, regulators, healthcare providers and patients [2, 3, 11, 12]. Issues with these devices can have life or death consequences [12, 13, 14, 15]. Therefore the patient end-user is the most significant stakeholder for these devices as it is their lives that are being saved. Still, the exact role of the patient as far as the security of these devices is unknown and requires further research [13, 14, 12,]. Therefore, the study of cybersecurity risk and threat communications in the era of the healthcare internet of things is prescient.

The following case study examined the research question: How are the cybersecurity risks and threats related to wireless implanted medical devices being communicated to patients who have or will have these devices implanted in their bodies? The case study provided timely, empirical data on how device manufacturers and medical professionals are guiding MIoT patients towards increased cybersecurity awareness and the ability to make sound decisions regarding cybersecurity risks and threats [15]. Because the field healthcare cybersecurity is new, and its policies and governance are still evolving [13], research about patient cybersecurity awareness supports the development of effective cybersecurity policies and procedures in healthcare.

The cybersecurity of wireless implanted medical devices is a relatively new field of interest [13], yet an important one. Not only has it been shown that there are substantial cybersecurity threats in the field of wireless medical device technology [16], but it is certain that as medical devices become increasingly connected and interoperable with other clinical systems, the cybersecurity threats to these devices will also increase [7]. Because of the rapid emergence of this field and its burgeoning complexity [12], the cybersecurity of wireless implanted medical devices provide fertile ground for research in many areas.

A few of the networked medical devices threatened by cyberattacks include Drug Delivery Systems (DDS), Biosensors, Neurostimulators and Cardiac implanted devices [15]. Pacemakers were chosen as the MIoT de-vice of interest for the case study to be discussed. Pace-makers manage heart rhythm disorders, such as bradycardia (when the heart beats too slowly) or arrhythmia (when the heart beats at an irregular rate). Pacemaker patients were ideal interconnected medical device patients to consider for several reasons.

The first reason was that all the major pacemaker manufacturers have wireless-enabled devices. What is more, these manufacturers also provide remote monitoring systems, which allow authorized personnel to access, monitor, and make changes to the devices from a distance. The most important reason, however, is that cardiac implanted device patients are more vulnerable to immediate, life-threatening harm than any other MIoT patient cohort is. Thus, it seemed reasonable to conclude the MIoT patients faced with the most severe consequences due to cybersecurity risks and threats might be inclined to have the most significant amount of cybersecurity awareness education and prove to be a model for all other MIOT patients.

Thematic Analysis was employed to identify and convey conceptual findings within this study. Thematic Analysis (TA) was used for its flexibility in performing data analysis in a variety of qualitative methods including case studies [17]. Through the application of TA, initial codes were created, which facilitated organizing the data into concepts and patterns. Those concepts and patterns were subsequently categorized, or grouped, into initial themes. The initial themes were then re-examined and reanalyzed, as suggested by [17] and [18], to arrive at five core themes, whose interpretation constituted the study's research findings [17].

## 4. DATA COLLECTION

The goal of the data collection was to provide the richest amount of salient data over a manageable length of time. Data collection was accomplished through the use of semi-structured interviews and observation [19]. The semi-structured interviews were held over the phone with participants from across the United States. The participants for this study were cardiac device specialists working in the U.S. possessing at least one year of experience working with cardiac implantable electronic device (CIED) patients who actively used home monitoring systems. Because the participants possessed distinct characteristics which made them rich sources of information for the study, purposive sampling was used to ensure all participants mirrored the target population [20]. The objective of the data collection was to speak with people who worked directly with CIED patients experiencing and undeniable need for cybersecurity awareness and education.

Table 1. Themes and Supporting Patterns

| Themes | Supporting Patterns |
|---|---|
| 1. Media Influence | 1.1 Negative Influence of the Media |
| | 1.2 Coping with Cybersecurity Concerns |
| | 1.3 Leveraging the Media |
| | 1.4 Cybersecurity Awareness Catalyst |
| 2. Risk/Benefit Analysis | 2.1 Risk of Hacking |
| | 2.2 Threat to Human Life |
| | 2.3 Risk/Benefit Equation |
| | 2.4 Maintaining Status Quo |
| 3. Non-Communication | 3.1 Non-Communication as a Standard Operating Procedure |
| | 3.2 Reactive vs. Proactive Communication |
| | 3.3 Patient's Right to Know |
| | 3.4 The Need for Transparency |
| 4. Need for Collaboration | 4.1 Cybersecurity Awareness and Perception |
| | 4.2 CyberAwareness Resources |
| | 4.3 Education for Patients and Providers |
| | 4.4 Cybersecurity Awareness Models |
| 5. Patient Care Dynamic | 5.1 Healthcare Provider's Role |
| | 5.2 Device Manufacturer's Role |
| | 5.3 Regulatory Environment |
| | 5.4 Device Management |
| | 5.5 Device Patient Characteristics |

Nearly all cardiac CIEDs released into the market possess remote monitoring features [21]. Likewise, all the major CIED manufacturers, i.e., Medtronic, Abbott (formerly St. Jude Medical), Boston Scientific, Biotronik, and Livanova offer in-home monitoring systems e.g. Carelink, Merlin@home, Latitude, Biotronik Home Monitoring, and Sorin SmartView, respectively. The combination of these two factors (in addition to a high-profile FDA cybersecurity device recall) made cardiac device specialists experienced with CIED patients using these in-home monitoring systems an ideal population for the study. The cardiac device specialists in this, most of whom held over a decade of applicable experience and served in the lead, supervisory, and management roles in their organizations tended to have very limited availability due to the nature of their work. Thirty-minute interviews were arranged to accommodate their varying and often very tight schedules.

## 5. DATA ANALYSIS

Thematic analysis was employed to identify and convey conceptual findings from this study. Thematic analysis is commonly used for its flexibility in performing data analysis in a variety of qualitative methods including case studies [17]. Thematic analysis was used in this case to create and codes that facilitated organizing the data collected into concepts and patterns, which could be categorized or grouped into initial themes. The initial themes were then re-examined and reanalyzed as suggested by [17] and [18], to arrive at themes, which could be used to interpret the data collected into meaningful research findings [17].

Each of the sixteen semi-structured interviews was transcribed verbatim. These transcripts became the dataset to which an initial set of thematic analysis codes were applied manually. The initial set of thematic analysis codes were derived directly from the research question to ensure, all subsequent coding and analysis were aligned with the research question. Thus, the initial codes were delineated as follows: HB (for how are); CB (cybersecurity); RB (risk/threats); DB (to WIMD devices); CM (communicated); PB (to patients). Additionally, the initial codes contained a separate, sequence of sub-codes, which identified the participant involved as well as the interview question answered at that time. This level of detailed coding created a set of focused codes derived from each participant’s complete set of interview responses. These initial and focused codes were imported into a data analysis software program (NVivo 11) for further organization and analysis.

The next level of thematic analysis coding, i.e. pattern coding and categorization, was performed with the help of the NVivo 11 data analysis software. Once the data had been drawn out (or reduced) through focused coding, the deeper analysis involved searching for patterns within the data and organizing those patterns into cogent data interpretation [17, 22]. A key advantage to using thematic analysis is that thematic analysis has quality and rigor built into the process. The process calls for the recursive examination and reconsideration of coding, categorization, and themes before, during and after they are analyzed and developed to confirm their validity against the data collected [17]. The themes derived from the data collected during the sixteen semi-structured interviews are illustrated in Table 1.

Additionally, Table 2 shows exactly which case studies were used as the primary data sources for each theme and theme element. It is important to point out that, at the level of theming, choices were made as to which participant’s input was delineative or depictive of a theme. For example, while every participant certainly spoke to the subject of cybersecurity awareness only those participant’s whose contribution aligned with the formation of a coherent pattern (such as the requirement for cybersecurity education) was tracked. Likewise, while every participant spoke of patient care, only three of the sixteen talked about the impact of regulations or a regulatory environment.

Table 2. Thematic Analysis Results with Total Number of Codes and Data Sources

| Themes and Supporting Patterns | Codes | Data Sources Cited (Cases 01-16) | | | | | | | | | | | | | | | |
|---|---|---|---|---|---|---|---|---|---|---|---|---|---|---|---|---|---|
| | | 01 | 02 | 03 | 04 | 05 | 06 | 07 | 08 | 09 | 10 | 11 | 12 | 13 | 14 | 15 | 16 |
| *1. Media Influence* | | | | | | | | | | | | | | | | | |
| 1.1 Negative Influence of the Media | 11 | X | X | X | X | X | X | | | X | | | | | | | X |
| 1.2 Coping with Cybersecurity Concerns | 6 | | X | X | | X | | | | | | | | | X | | X |
| 1.3 Leveraging the Media | 1 | | X | | | | | | | | | | | | | | |
| 1.4 Cybersecurity Awareness Catalyst | 5 | X | X | | | X | | | | X | | | | X | | | |
| *Total* | 23 | X | X | X | X | X | X | | | X | | | | X | X | | X |
| *2. Risk/Benefit Analysis* | | | | | | | | | | | | | | | | | |
| 2.1 Risk of Hacking | 12 | X | X | | X | X | | | | | | | X | | | X | |
| 2.2 Threat to Human Life | 11 | X | X | X | X | X | X | X | | | X | | | | | | |
| 2.3 Risk/Benefit Equation | 18 | | X | X | X | X | X | X | | X | | | | | X | | X |
| 2.4 Maintaining Status Quo | 5 | | X | X | X | | X | | X | | | | | | | | |
| *Total* | 46 | X | X | X | X | X | X | X | X | X | X | | X | | X | X | X |
| *3. Non-Communication* | | | | | | | | | | | | | | | | | |
| 3.1 Non-Communication | 20 | X | X | X | X | X | X | X | X | | X | X | X | | X | X | X |
| 3.2 Reactive vs. Proactive | 16 | X | X | X | | X | X | X | X | X | | X | | | | | X |
| 3.3 Patient's Right to Know | 15 | X | X | X | | X | | X | X | X | | | | | X | | |
| 3.4 Need for Transparency | 12 | X | X | | X | X | X | | | | X | X | | | X | X | |
| *Total* | 63 | X | X | X | X | X | X | X | X | X | X | X | X | | X | X | X |
| *4. Need for Collaboration* | | | | | | | | | | | | | | | | | |
| 4.1 Cyber Awareness and Perception | 21 | X | X | X | X | X | X | X | X | | X | | X | | | X | X |
| 4.2 CyberAwareness Resources | 27 | X | X | X | X | X | X | X | X | X | X | X | X | X | X | X | X |
| 4.3 Education for Patients and Providers | 39 | X | X | X | | X | X | X | | X | | X | X | X | X | X | X |
| 4.4 Cybersecurity Awareness Models | 20 | X | X | X | | X | | | X | | X | X | X | | | X | |
| *Total* | 107 | X | X | X | X | X | X | X | X | X | X | X | X | X | X | X | X |
| *5. Patient Care Dynamic* | | | | | | | | | | | | | | | | | |
| 5.1 Healthcare Provider's Role | 41 | X | X | X | | X | X | X | | | X | X | X | | | | X |
| 5.2 Device Manufacturer's Role | 38 | X | X | X | X | X | X | | X | X | X | X | X | X | X | | |
| 5.3 Regulatory Environment | 4 | X | X | | | X | | | | | | | | | | | |
| 5.4 Device Management | 29 | X | X | | X | X | X | X | X | X | | X | X | X | | | |
| 5.5 Device Patient Characteristics | 48 | X | X | X | X | X | X | X | X | X | X | X | X | X | X | X | X |
| *Total* | 160 | X | X | X | X | X | X | X | X | X | X | X | X | X | X | X | X |

## 6. RESULTS

The sixteen interviews in this multiple case study constituted a little more than eight hours of conversations held over 31 days. The transcription process, on the other hand, took over 60 hours and yielded 85 pages of verbatim transcripts. This was followed by well over 100 hours of coding and analysis over the course of another 51 days. The results present a compelling picture of how wireless implanted medical device patients (cardiac implanted medical device patients using home monitoring systems) are provided with information about the cybersecurity risks and threats to the devices implanted in their bodies. Figure 1. provides a cluster analysis of the top 50 words within the five thematic analysis themes.

Typically, a patient's first source of information, for better or worse, is the media. Sparked by what they have seen in newspapers, radio, TV, movies and social media, patients contact their healthcare provider to gain answers to their cybersecurity concerns. Two things are interesting to note. The first being every manufacturers' website has information available to the public about medical device cybersecurity. However, the second thing to note is when case study participants were asked, what resources and materials are available to help device patients with cybersecurity awareness most of them answered, "Nothing."

This three-dimensional diagram clusters similar terms together and spreads different terms apart demonstrating how far the word "media" stands out within the keywords of the study's results.

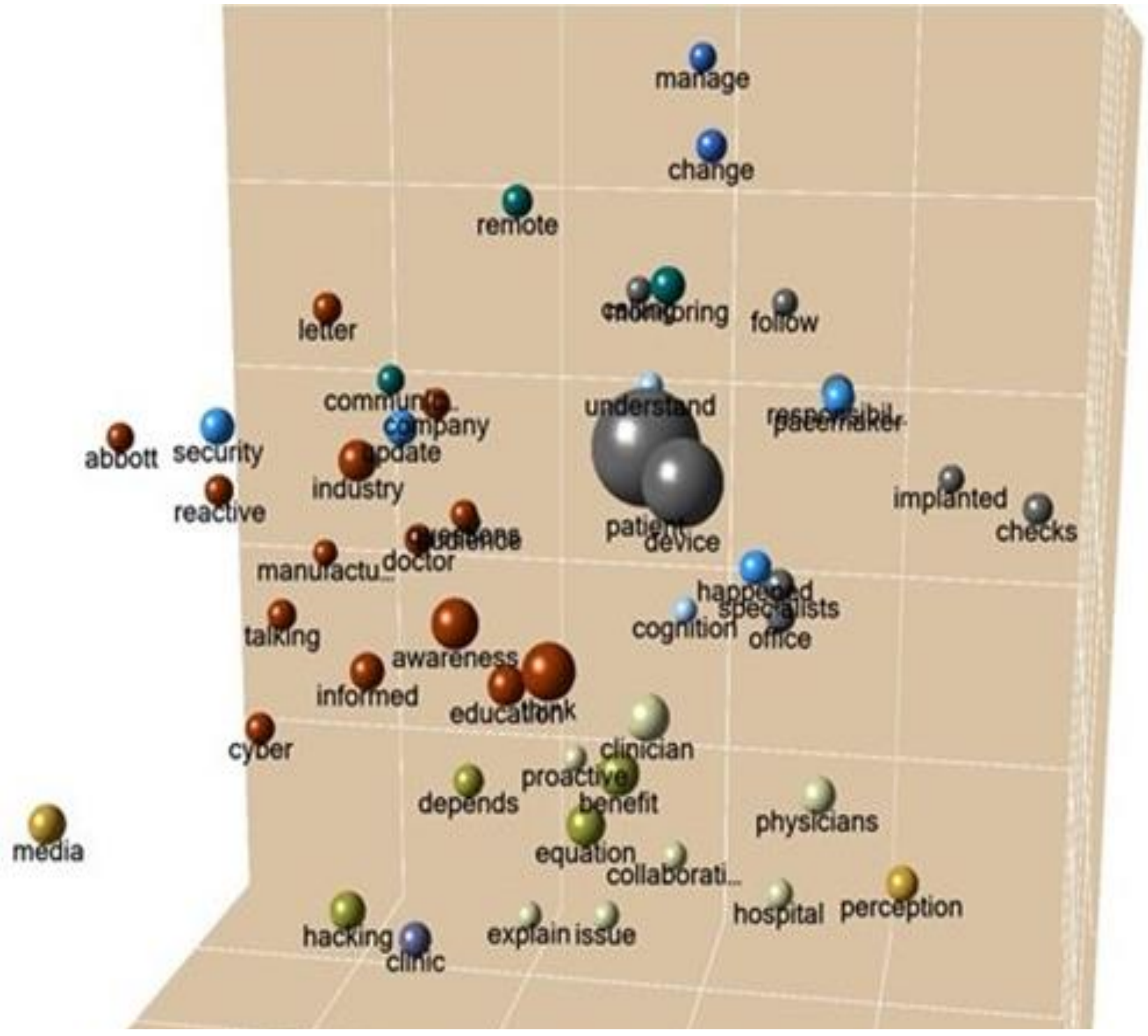

Figure 1. Cluster analysis of the top 50 words within the five thematic analysis themes.

The explanation offered by several study participants was that only younger patients or technically well-informed people would consult a manufacturer's website for information before seeking the advice and assurance of their healthcare professional. Moreover, it was widely reported in the study patients felt most assured after speaking with their cardiologist or electrophysiologist even more so than speaking with experienced and highly trained nurses or trained and certified device technician and specialists or even the device manufacturers' representatives.

Nevertheless, the study established cardiac device nurses and cardiac device technicians and specialist not only take the initiative to train and educate themselves about CIED cybersecurity (with assistance from the manufacturer's representatives, websites and the use of other resources) but also, for the most part, it is these nurses, and device technicians and specialist who (either single-handedly or as a group) create, develop and disseminate the content needed to provide the cybersecurity training and awareness patients receive.

It was found. However, the method of operation for this process is extremely reactive rather than proactive. The overwhelming preference, as practiced to date, is to respond to patient concerns on demand. This is because, the consensus opinions expressed by participants was that no harm has ever come to a CIED patient through cybersecurity threats—therefore the risk to these devices and the patient's health are small to be worth scaring the patient over.

Finally, in speaking of the patient, the second gap in the literature was the question of the patient's role in MIoT cybersecurity. The most important role the patient can play is to understand their device, understand what exactly their device does, and understand all the systems which support their device. Thus, it was agreed, the patient's role

## 7. CONCLUSION

The thematic analysis applied to this study yielded five themes across sixteen cases. Those five themes are: the influence of the media on patient cybersecurity awareness; the need for risk/benefit analysis at all levels of patient interaction; the culture of non-communication in the healthcare and medical device industry; the need for collaboration and education among manufacturers, healthcare providers and patients; and the obstacles and challenges present in all phases of patient care.

The first goal was to understand how patients were receiving information about the cybersecurity risks and threats to the MIoT device implanted in their bodies. It was found their primary and most consistent source of information came from the media, not from the healthcare providers or device manufacturers. It was shown; because of the everyday risk vs. benefit calculations made on the part of providers and industry, cyber-security communications to patients were typically muted and reactive. There was information available for those who knew where to look or what to ask. Otherwise, the consensus among participants was not to bring up the subject of cybersecurity until they were asked.

The second goal of the study was to understand the device patient's role. Research on the technical aspects of a cybersecurity attack and the defense still dominates the literature. While there is a growing body of literature concerning cybersecurity in healthcare operations, there is still little research on the role of the patient in MIoT cybersecurity. This study confirmed MIoT patients have a significant role in their device cybersecurity as recounted by a purposeful sampling of healthcare professionals working closest with these patients. In the estimation of these professionals, the patient must first understand the device they have in their bodies. Furthermore, the patient needs to understand what their interconnected medical device does, how and why the device communicates with the outside world and understands the ramifications of the MIoT communications as far as the patient is concerned. There is also a definite, undeniable responsibility of the patient to remain engaged and responsive to their health care team in every phase of device management (including cybersecurity).

It is the responsibility of the patient (or whoever must support the patient) to arm themselves with reliable data regarding the risks and threats to their interconnected medical device. Patients have the responsibility, for their own sake, to learn enough about the device in their body so that when they receive information from their manufacturer, healthcare provider, or even the media, they can actively participate in the decision-making process about the next steps to be taken in their best interest. Being active means staying curious, asking questions, communicating with the manufacturer's teams, their healthcare team, their support networks and demonstrating they are an active participant in their health and welfare in an increasingly interconnected world.

### 7.1. Future Study

Further research is recommended in factors that govern and influence patient cybersecurity perceptions such as the role and influence of the media, family relationships, or other patient support systems. As patient demographics were heavily cited as a barrier to patient communication, research on optimal approaches to MIoT cybersecurity [28] awareness within an aging demo-graphic could be useful. Time would not permit a broad analysis of interconnected medical devices. Thus, studies specifically about patient cybersecurity risk and threat communications regarding other interconnected medical devices such as insulin pumps, neurostimulators, cochlear implants and the like may be useful. More significantly, because the study is confined to the United States a similar study in other countries (or even globally) could offer value and additional insight. An-other study opportunity would be to explore the perspectives of physicians or the viewpoints from the MIoT device industry. Last, there is the opportunity to continue research with MIoT patients as far as their lived experience in managing the risks and threats they perceive to their device.

## AUTHOR

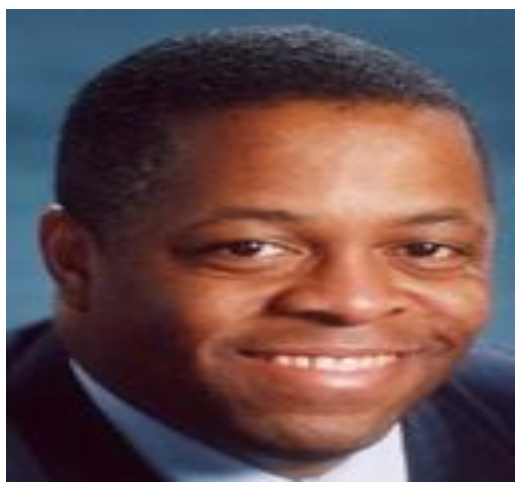

**Dr. George W. Jackson, Jr.** has nearly 30 years of experience in Information Technology, Information Security and IT Project Management. A seasoned business and technology expert possessing an MBA and IT Management and professional certifications in Project Management, Information Security and Healthcare Information Security and Privacy. He has earned a Ph.D. in Information Technology, Information Assurance and Cybersecurity from Capella University.

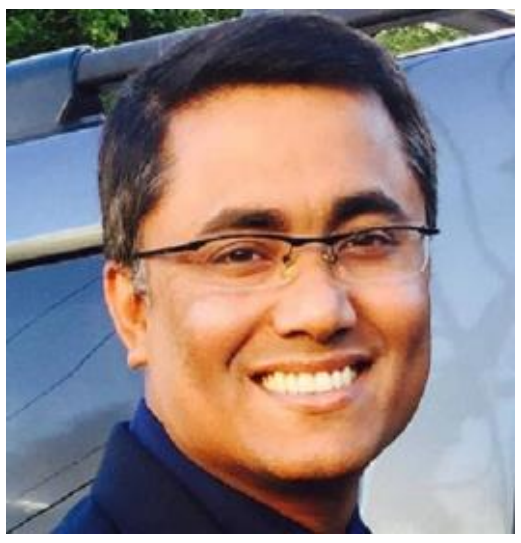

**Dr. Shawon S. M. Rahman** is an Associate Professor in the Department of Computer Science and Engineering at the University of Hawaii Hilo. His research interests include software engineering education, information assurance and security, web accessibility, cloud computing, and software testing and quality assurance. He has published over 120 peer-reviewed papers. Dr. Rahman is serving as the Member-at-large and Academic Advocate: Information Systems Audit and Control Association (ISACA) at the University of Hawai‘i at Hilo and Academic Advocate of the IBM Academic Initiative. He is an active member of many professional organizations including IEEE, ACM, ASEE, ASQ, and UPE.